\documentclass[referee,a4paper,12pt,traditabstract]{swsc} 
\usepackage{graphicx}
\usepackage{txfonts}
\usepackage{subfigure}
\usepackage{epstopdf}
\usepackage{lineno}
\usepackage[authoryear,round]{natbib}
\usepackage[dvipdfmx]{hyperref} 
\usepackage{url}
\hypersetup{colorlinks=true,citecolor=cyan,urlcolor=cyan,linkcolor=blue}

\begin{document}
\title{Tracking of magnetic flux concentrations over a five-day observation, and an insight into surface magnetic flux transport}
\titlerunning{Tracking of magnetic concentrations}
\authorrunning{Iida}
\author{Y. Iida \inst{1} }
\institute{Institute of Space and Astronautical Science, Japan Aerospace Exploration Agency,
              3-1-1 Yoshinodai, Chuo, Sagamihara, Kanagawa 252-5210, Japan\\
              \email{\href{mailto:iida@solar.isas.jaxa.jp}{iida@solar.isas.jaxa.jp}}}

   \abstract
{The solar dynamo problem is the question of how the cyclic variation in the solar magnetic field is maintained.
One of the important processes is the transport of magnetic flux by surface convection.
To reveal this process, the dependence of the squared displacement of magnetic flux concentrations upon the elapsed time is investigated in this paper via a feature-recognition technique and a continual five-day magnetogram.
This represents the longest time scale over which a satellite observation has ever been performed for this problem.
The dependence is found to follow a power-law and differ significantly from that of diffusion transport.
Furthermore there is a change in the behavior at a spatial scale of $10^{3.8}$ km.
A super-diffusion behavior with an index of $1.4$ is found at smaller scales, while changing to a sub-diffusion behavior with an index of $0.6$ on larger ones.
We interpret this difference in the transport regime as coming from the network-flow pattern.}
{}        
   \maketitle
\section{Introduction}

Magnetic fields on the solar surface are a fundamental driver of solar activity, with its effects upon the Sun-Earth system. 
The solar dynamo problem, whose main manifestation is the 11-year cycle in solar activity, and the 22-year cycle in the solar magnetic field, is of great concern to solar physics and geoscience.

The flux-transport model is one of the plausible scenarios to account for this cyclic variation.
The basic theory behind the model was proposed by Babcock and Leighton \citep{bab1961, lei1964}.
One important process in this model is poleward transport of magnetic flux at the solar surface.
Sunspots appear near 30 degrees latitude with a certain lean angle from the longitudinal direction and are dispersed within one or two months; in line-of-sight observations of the photospheric magnetic field, the Sun reveals a patchy structure at small scales ($\leq 10^3$ km), called magnetic flux concentration.
The magnetic field can strengthen until it becomes an equipartition field, at which stage the magnetic energy density is equal to that of the kinetic energy; it may strengthen even further by the process of convective collapse.
However, most of the magnetic field remains weaker than the equipartition field, and the motion of the concentration is thought to be dominated by that of the plasma.
Hence plasma flows play an important role in magnetic field transport.
These flows can be divided into two categories, global-scale and small-scale flows.

There are two kinds of global-scale flows at the solar surface.
One is the differential rotation, which is the variation in the rotational velocity as a function of latitude and expanding the surface magnetic region along the longitudinal direction.
The other one is the meridional flow, which is a flow pattern from the equator to the pole in each hemisphere. 
It transports the magnetic concentration to the pole and is important for the length of the solar cycle.
The amplitude of the meridional flow is on the order of $10$ m s$^{-1}$ \citep{bec2002,wan1989}; it takes $\thicksim$ 1 year for the magnetic flux concentrations to travel the distance of the solar radius ($\thicksim 7.0 \times 10^5$ km).
However, if only these global-scale flows were considered, then all the flux of positive and negative polarities in the sunspots would reach the pole and there would be no polarity reversal, e.g., no cyclic variation in the magnetic field.

The surface convective flow plays an important role here.
Convective flow patterns at the solar surface are multiscale \citep{nor2009}.
There are three kinds of convections on the solar surface; the smallest but strongest convective flow pattern is called granulation.
This can be observed directly in the intensity images of photospheric continuum.
The spatial scale of these granules is $\thicksim 10^3$ km, and their amplitude is $\thicksim 1$ km s$^{-1}$, which is larger than that of the meridional flow by two orders of magnitude.
Mesogranulation occurs on a larger scale than granulation \cite{nov1981} and has a spatial scale of $\thicksim 7 \times 10^3$ km and a horizontal velocity of $500$ m s$^{-1}$.
However, it is difficult to see the significant signature corresponding to mesogranulation. 
\cite{yel2011} investigated relationships between horizontal flow pattern from intensity images and magnetic field by the use of high spatial and temporal datasets obtained by the Imaging Magnetograph eXperiment (IMaX). 
Although there was a high correlation between the horizontal flow field and motion of magnetic concentrations, they found no significant scales within the range of $< 10$ Mm in probability density functions of footpoint separation distance. 
\cite{ber2013} investigated length scale of void structure of line-of-sight magnetic field but they found no preferred scales of organization in the range of $2 \-- 10$ Mm and suggested that the multiscale nature of flows on the solar surface masks a mesogranular scale.
The last and largest one type of solar convective flow is supergranulation.
It has a spatial size of $\thicksim 1.6 \times 10^4$ km, and a typical horizontal speed of $300 \-- 500$ m s$^{-1}$ (See \cite{rie2010b} and references therein).

Because these surface convections are much faster than the meridional flow, the magnetic flux concentration rapidly prevails during transport to the pole.
Considering the inclination of sunspots and anti-polarity between the hemispheres, this prevails across the hemisphere, resulting in imbalances of the cancellation occurrence between the polarities in each hemisphere. 
Hence, the flux to the pole is nonzero and changes in the magnetic polarity takes place.

\cite{wan1989} investigated this model by using numerical simulation.
They calculated the time evolution of the surface magnetic field by changing the meridional flow speed and the diffusion coefficient for smaller-scale convections.
The simulation with $10\--20$ m s$^{-1}$ as meridional flow and $600$ km$^2$ s$^{-1}$ as convective diffusion was found to produce the results that are the most consistent with the observations.

However, recent observations reveal that the magnetic field transport has a character that is different from the normal diffusion regime.
The relationship between the elapsed time and the squared displacement of the magnetic flux concentration is therefore investigated there because their proportionality is a necessary condition for the diffusion transport. 
The critical difference in the magnetic field observation between Leighton's era and the present is that the tracking of each magnetic flux concentration is possible now, allowing us to directly investigate the magnetic flux transport.
The tracking of the patches requires a stable spatial and temporal resolution better than $10^3$ km and $1$ minute respectively, which could not be accomplished in Leighton's era.
\cite{abr2011} summarize the recent situation and the super-diffusion scaling with an index varying from 1.27 to 1.67 is reported in recent work based on ground observation and satellite observations \citep{hag1999b,law2001,gia2014,car2015}.

However, there is still a large gap between the spatial scale of tracking and the global scale. 
This is because of limitations in the observation period.
Previous papers have mainly been devoted to time scales below one day. 
We need longer observations to investigate transport over larger scales.

Here, we investigate the relationship between the elapsed time and square of the displacement of the magnetic concentration in the longer time scale.
Recent developments in technology have facilitated this kind of the investigation in two ways; first, feature-recognition and tracking techniques of the magnetic flux concentration have advanced over the last few decades \citep{hag1999b,def2007,par2009,tho2011}.
The analysis of huge amounts of events is needed for investigating solar surface transport.
This is because the magnetic elements are much smaller than the global scale, and hence, the statistical character is crucial; feature-recognition techniques are plausible solutions to this difficulty.
Secondly, continual and uniform magnetogram data are now being provided by the satellites.
The uniformity of the data set is important for the feature-recognition method.
Moreover, while the maximum ground-observational period is limited by the Earth's rotation, satellite observation is free from this limitation and is thus preferable for this kind of study.

In section 2, we briefly summarize the basic concept connecting the global transport regime and the motion of each element in 1-D random-walk modeling.
Descriptions of the observational data and tracking algorithm are shown in sections 3 and 4, respectively.
The result is shown in section 5. 
Section 6 is devoted to a comparison between the present data and previous observations and theoretical requirements.

\section{Global transport regime and motion of each element}

In this work we find that the data set of magnetic field concentration exhibit properties of non-Fickian diffusion. 
We will focus on analyzing the relation between mean squared displacements and time and will not explore concept related to the fractal dimensionality of the space.
Therefore to introduce the main concepts we present here the case of the 1-D random walk model leading to normal diffusion.

A 1-D random walk is defined as a motion that has constant travel distance during one walk and constant waiting time between walks.
Fig.1 shows a schematic example of a 1-D random walk.
The expected position of the particle is clearly $\langle x(t) \rangle =0$ at any time under random motion.
However, the expected dispersion of the particle position, e.g., the squared displacement from $t=0$, increases with time and is calculated as
\begin{equation}
\langle x^2(t) \rangle =\left( \frac{\delta^2}{\tau} \right) t,
\end{equation}
where $\tau$ and $\delta$ are the temporal and spatial scales for one jump.
This result shows that the squared displacement is proportional to the elapsed time in the random walk of a particle. 
The analytical solution of the diffusion transport has the same character.
The diffusion transport of the praticle density in 1-D is written as
\begin{equation}
\frac{\partial n(x,t)}{\partial t} = D \, \nabla^2 n(x,t),
\end{equation}
where $n(x,t)$ is particle density at position of x and time of t and $D$ is a diffusion coefficient.
The expectation of $\langle x^2(t) \rangle$ can be calculated as 
\begin{equation}
\langle x^2(t) \rangle =\int^{\infty}_{-\infty} \, x^2 n(x,t) \, dx = 2Dt.
\end{equation}
Again, we see that the expectation value of the squared displacement is proportional to the elapsed time.
Diffusion transport gives the same expectation value of the dispersion as the random walk particle with a condition $D=\delta^2 / 2\tau$.
Note that this proportionality is a necessary condition for the description by diffusion transport on a global scale.

Another simple case is ballistic transport.
In this case, the particle motion is written as $dx(t)/dt=v_0$ and hence $ \langle x^2(t) \rangle =v_0^2 t^2$.
Here, the squared displacement is proportional to the square of the elapsed time.

This concept, the relationship between the elapsed time and the squared displacement, can be expanded to the more general transport regime, which is called non-Fickian diffusion \citep{bal1988}, e.g. anomalous diffusion.
Fig.2 shows a schematic picture of this concept.
The sub-diffusion regime is where the power-law index is less than $1$, meaning that the spreading of the particles slows down as compared to what would happen in a diffusive regime.
A typical mechanism of this regime is trapping with random walks.
When there are trapping points, some of the particles are captured and the average spreading velocity becomes slower than in the diffusion regime. 
In contrast, there is a super-diffusive regime between diffusion and ballistic motion.
The power-law index then has a value between $1$ and $2$, indicating faster spread than under ballistic motion.
A typical case of super-diffusion is the L\'evy flight, wherein the velocity of the particle has a power-law distribution rather than a Gaussian one.
Above ballistic motion, we find the hyper-diffusion regime.
It has a power-law index larger than $2$ and the pervading speed increases as it spreads.
A self-avoiding random walk in which the paths of the elements never interact with each other is a typical case of the hyper-diffusion.
However, this is not the case for solar surface transport.
There is another idea to explain the non-Fickian diffusion mechanism.
\cite{sch1990} investigated motions of magnetic concentrations in the core and surroundings of active regions, and found different amplitudes for the diffusion coefficient.
They discussed that the scatter in the amplitudes may be caused by differences in step length because there were no significant difference in velocity distributions.
If the diffusion coefficient varies with space, then the diffusive transport may be globally anomalous, while being diffusive within each region.

Unlike the ballistic and diffusion cases, a fractal differential equation is necessary to include these non-Fickian processes.
Please see \cite{bak2008} for a more systematic and mathematical treatment.

Thus, we can see that the relationship between the elapsed time and squared displacement contains important information about the form of the global transport, although it is only a necessary condition.
We investigate this relationship in the transport of magnetic flux concentration on the solar surface.

\section{Instruments and Data}

For global-scale transport, the most important point is the duration of the observation.
Because the observational duration of a ground observatory is limited by the Earth's rotation, it is not suitable for this kind of analysis.
We use satellite observations in this study.
The second important point is spatial resolution.
It is still difficult to detect small magnetic flux concentrations from recent satellite data although their spatial resolution has been improved. 
Higher spatial resolution makes feature recognition much easier.
For these reasons, we select the magnetograms obtained by the Solar Optical Telescope (SOT) onboard the Hinode satellite \citep{kos2007}.

The Hinode satellite was launched in September 2006.
SOT is one of the three telescopes onboard the Hinode. 
It observes the Sun in the visible spectrum and provides spectropolarimetric data, from which we can calculate the magnetic information on the solar surface.
The SOT has a high spatial resolution of $0.2 \-- 0.3''$, although it cannot cover the entire Sun.
There are two instruments on SOT: a filtergram (FG) and a spectropolarimeter (SP).
We use the magnetogram obtained by the FG in this study.
The FG consists of two filtergram imagers: the broadband filtergram imager (BFI) and the narrowband filtergram imager (NFI).
NFI obtains the Stokes-polarization signal in several lines, and the magnetic field information on the solar surface is calculated from the polarization signal based on the Zeeman effect.
We use the magnetograms of the Na I 589.6nm absorption line obtained by NFI. 

The longest observation for which magnetogram data is available, taken between September 2006 and June 2014, is used in this study.
SOT observed the quiet region from 10:24 UT on December 30$^{\rm th}$, 2008 to January 5$^{th}$, 2009.
The total duration of the observation was 115 hours 13 minutes,which is $\thicksim 5$days.
The temporal cadence and field of view was limited to $\thicksim$5 min and $\thicksim 110''\times110'' = 6 \times10^9$ km$^2$, respectively.
1,642 magnetograms were obtained in total.
The pixel scale was $0.16'' \thicksim 120$ km.

Data corrections are needed before the analysis.
We performed the same correction as \cite{iid2012}.
Dark- and flat-field corrections of the CCD camera are done with fg$\_$prep.pro, contained in the SolarSoftWare (SSW) package.
The column-wise median-offset of the CCD camera is known and the median value of each column is subtracted from all pixels in the same column \citep{lam2010}.
Due to solar rotation, the position of the observing center moves from solar coordinates of ($-520.9''$,$-9.3''$) to ($695.9''$,$-6.9''$).  
All images are de-rotated to 8:01UT January 2$^{nd}$, 2009, which is the middle of the observing period, by the procedure drot$\_$map.pro in the SSW package.
Although the pointing stability of Hinode is excellent, thanks to a correlation tracker, a small discrepancy still exists.
We investigate this discrepancy by determining a correlation between the consecutive images.
In this long data set, the discrepancy reaches a value of $15'' \thicksim 10^4$ km.
Only the region observed over the whole period is analyzed, which has an angular size of $94.9'' \times 91.4''$. 
We need the conversion factor between the polarization signal and the actual magnetic field, although a preliminary conversion is done onboard.
This conversion factor is derived from the linear fitting between the circular polarization (CP) of SOT/NFI and the line-of-sight magnetic field derived from the Michelson Doppler Imager (MDI) onboard the Solar and Heliospheric Observatory (SoHO). 
The factor is derived as 1.83 G DN$^{-1}$.
Fig.3 shows an example of the magnetogram after all the above corrections.

\section{Feature-tracking Algorithm}

We use the same algorithm for the recognition and tracking of the magnetic concentration as \cite{iid2012}.
We briefly summarize their method here.

In the recognition process, we use the clumping method by \cite{par2009}, where we employ a signal threshold and the pixels above it are recognized as valid pixels.
The threshold is determined in each magnetogram by fitting the histogram of the magnetic field strength with a Gaussian function.
For our purposes, we employ two $\sigma$ as the threshold.
The $\sigma$ varies around $5$G in this data set , which is typical value for quiet Sun regions, and hence we set $\thicksim 10$ G as a magnetic field strength threshold in this study.
The size threshold, which is used as the smallest size of the recognized concentration, is set to be 81 pixels, or $\thicksim 5 \times 10^5$ km$^2$, corresponding to the granular size.

After the recognition, the concentrations between the consecutive magnetograms are tied by comparing the spatial overlaps, which are checked with an extra $5$ pixel margin. This is necessary because of the 5-minute interval between the magnetograms.
In some cases, more than one concentration overlaps with that in the previous magnetogram.
The concentration that has the most similar flux content is related to the previous concentration in these cases.
Note also that tracking is done from the concentration with a larger flux content.
This is because the smaller concentrations may have a tendency to disappear or stay smaller than concentrations with a larger flux content.
With the tracking method explained above, the concentration with the largest flux content survives after coalescence and splitting.

Fig.4 shows the schematic picture of the method.
In this case, a concentration of negative polarity starts from the bottom left of the figure.
It moves and splits into two concentrations at a certain time.
During this splitting, the concentration with larger flux content in the latter magnetogram is treated as the same as the originally tracked concentration.
The concentration with smaller flux content is treated as one that has newly emerged.
After the splitting, the originally tracked concentration collides with one having the opposite polarity.
In this case, three concentrations exist in the tracking method, which are shown by the orange arrows in Fig.4.
The lifetime and displacement from birth to death for each concentration are recorded.

After tracking all concentrations, the elapsed time and squared displacement are obtained for each. 
Fig.5 summarizes all possible cases of the birth and death of the magnetic flux concentrations in the analysis.
The concentrations surrounded by the green square are the newly emerged or vanished ones in the left and right columns respectively.

\section{Results}

In total, 21823 positive-polarity and 19544 negative-polarity concentrations are tracked.

First, we investigated the typical lifetime of the magnetic concentrations.   
The average lifetimes are estimated to be 22.5 and 23.9 min for the positive- and negative-polarity concentrations, respectively.
Further, the lifetime distribution is investigated.
Fig.6 shows the number histograms of the lifetimes, using logarithmic axes. 
The thin solid and dashed histograms respectively correspond to positive and negative concentration. 
The thick solid histogram shows the total of both polarities.

Next, we investigate the relationship between the elapsed time and the squared displacement of the magnetic concentrations, as displayed in Fig.7.
The horizontal axis is the lifetime from birth to death of the concentrations and the vertical axis is the squared displacement.
The bins are set [$10^x$, $10^{x+0.1}$].
Diamonds indicate the average squared displacement in each bin; bars indicate the standard deviations in each bin as an error of this analysis.
We see the change in the dependence around scales of $2 \times 10^4$ s (corresponding to a squared displacement of $10^{7.5}$ km$^2$ , or a length scale of $10^{3.8}$ km, and hence we fit the results into two domains.
The dashed and dotted lines show the results of fitting in the range below and above $2 \times 10^4$ s, respectively.
The dashed one has a power-law index of $1.4\pm0.1$ (super-diffusion scaling) and the dotted one has a power-law index of $0.6\pm0.2$ (sub-diffusion scaling).

Recent papers have investigated the separation distance of the paired magnetic flux concentrations \citep{rep2012, gia2014b}.
Here, we perform the same analysis on our data set.
Fig.8 shows the dependence of the squared separation distance on the elapsed time, again with logarithmic axes.
Diamonds indicate the averaged squared separation in each bin. 
We see a linear relationship in this plot and hence fit the result with a line with the range below 10$^5$ s.
The result of the fitting is shown by the dashed line.
We obtain the power-law index of $1.46$ with the fitting.
This value is consistent with recent papers.
\cite{rep2012} reported an index of $1.5$ over spatial scales from $10^{1.2}$ km to $10^{2.5}$ km  and temporal scales from $10$ s to $400$ s. 
\cite{gia2014b} reported $1.55$ for a quiet region over spatial scales from $10^{2}$ km to $10^{3.6}$ km and temporal scales from $10$ s to $10^4$ s.
Our result extends their result to larger temporal and spatial scales.

\section{Discussion}

We have investigated the relationship between the elapsed time and the squared displacement of magnetic flux concentrations over the longest observation by the Hinode satellite and found a different behavior above and below the scale of $10^{3.8}$ km. 
On short temporal and spatial scales, the super-diffusion regime has an index of $1.4$.
This scaling is consistent with previous studies \citep{law2001,abr2011,gia2014}, where the power-law index varies from 1.27 to 1.67.

On the other hand, we have newly found in the quiet Sun that the scaling becomes sub-diffusive above the scale of $10^{3.8}$ km.
\cite{law1993} reported sub-diffusion scaling in the active region for temporal scales longer than half a day, and spatial scales larger than $6 \times 10^3$ km.
These scales are similar to the ones found here.

What causes this change of the scaling near $10^{3.8}$ km?
We suggest the supergranulation as a plausible candidate; although the strongest convective flow pattern is a granular flow as mentioned, the magnetic field is also forms magnetic network due to being transported by supergranular flow, whose mechanism remains unknown.
The typical scale of the network field is $\thicksim 10^{4}$ km, which is slightly larger than that of the change found in this paper.
Our interpretation is as follows: inside the network boundary, the magnetic concentration is affected not only by the granular flow patterns but also by the network flow pattern.
The granular flow pattern is expected to cause a diffusive-like transport of the magnetic field because its characteristic size is smaller than that of the network.
On the other hand, the network flow pattern is expected to transport magnetic field ballistically below the scale of $10^4$ km because its cell size is larger; hence, it is natural to assume ballistic transport by the network flow pattern.
With diffusive transport by the granulation and ballistic transport by the network flow pattern, we expect an intermediate scaling between them, namely super-diffusive transport.
In a larger scale than the network field, it is expected that the magnetic concentration will reach the conjunction point of the network.
At this point, the magnetic field will be trapped by the network flow.
This trapping is expected to result in the sub-diffusion scaling above $10^4$ km.
To support the hypothesis that the magnetic field is trapped by network downward flows, dopplergrams can be used to identify the co-spatiality of the stagnation points and intense downflows in the future study.
Fig.9 shows a schematic view of this speculation. 
Numerical simulation is also a powerful tool to resolve this matter.

A recent paper by \cite{gia2014} found a change in the power-law index from $1.34$ below the scale of granules ($1.5 \times 10^3$ km) to $1.25$ above it.
The spatial scale is smaller by one order of magnitude than the findings in this paper and it remains in the super-diffusion region.
The reason we cannot see this may be that the temporal resolution is poorer in the dataset of this study than in theirs.
The time interval of our dataset is 5 min, which is insufficient for investigating granular scales of $\thicksim 10$ min.
Thus, it is difficult to see the change in the dependence around the granular scale.
Why does the super-diffusion dependence continue at temporal scales longer than granulation?
We believe the answer lies in the different balance between the magnetic and kinetic energies of convection.
The conjunction point of the network is known as a place where strong magnetic fields exist; it is reasonable to expect such magnetic fields to prevent convective flow from prevailing.
On the other hand, the magnetic field at the edge of a granule is weaker and may not be able to stop the convective flow.
We believe that this difference may cause the difference in their behaviors.

Next, we will discuss the inconsistency with the theoretical modeling. 
The requirement for diffusion coefficient from kinetic dynamo simulation is $600$ km$^2$ s$^{-1}$ \citep{wan1991}; in previous papers, this value has been reached approximately at the network scale ($\thicksim 10^4$ km).
If the dependence of squared travel distance on time has sub-diffusion scaling on longer scales, the effective diffusion coefficient should decrease at such large scales.
Considering the 1.5 or 2 orders of spatial difference between the network and the solar global scale, the effective diffusion coefficient becomes less than the requirement by one order of the magnitude from the present result.

Although the sub-diffusion scaling obtained in this study covers only one order of magnitude, the investigation of longer time scales is difficult because Hinode has a geocentric orbit and the observational duration is limited by the Sun's rotation. 
We need a heliocentric satellite for longer time scales.
However, the accuracy of the result can be improved with the accumulation of the dataset.
The dependences among quantities such as latitude, longitude, and magnetic flux amount are an interesting question for future research.

At the last part of the discussion, we shall suggest plausible processes to fill the gap between the measured value of the diffusion coefficient and that required by numerical simulation.
We assumed rigid particles without interactions and evaluated the diffusion coefficient of the horizontal motion on the plane, e.g., the photosphere. 
Some processes are not considered in this modeling.
One is the merging and splitting of the magnetic flux concentrations.
When magnetic concentrations merge together and then split, the magnetic flux is transported.
Hence, the merging and consequent splitting increases the transport coefficient.
Because recent studies reveal that merging and splitting of magnetic flux concentrations frequently occur on the actual solar surface \citep{lam2008, lam2010,iid2012, gos2014}, we expect that this process is dominant in global transport.
\cite{iid2015} investigated the same dataset with this study and found the frequent merging and splitting of magnetic concentrations with a time scale of $\thicksim 30$ min.
Other processes that are not considered in this study are cancellation and emergence. 
The magnetic field is transported in the vertical direction by these processes, making it possible for the magnetic field to go through a conjunction point, increasing the effective transport coefficient.
Again \cite{iid2015} investigated cancellation events in this dataset but found them less frequent than merging and splitting.
However, the dependence of cancellation occurrence on magnetic flux has a power-law index of $-2.48 \pm 0.26$.
The power-law index less than $-2$ implies that smaller cancellation is important in terms of total amount of magnetic flux transport.
Hence there is a possibility that cancellation events smaller than those investigated in \cite{iid2015} transport significant magnetic flux.
We expect either or both of these processes to fill the gap between the theoretical requirement and the result in this study.  
However, there have been no theoretical models taking these processes into the account so far.
The matter of including them in the global transport equation is left for future work.

\begin{acknowledgements}
First of all, we would like to express my thanks to the Hinode team for providing the continuous observational data.
Hinode is a Japanese mission developed and launched by ISAS/JAXA, collaborating with NAOJ as a domestic partner, NASA and STFC (UK) as international partners. Scientific operation of the Hinode mission is conducted by the Hinode science team organized at ISAS/JAXA. This team mainly consists of scientists from institutes in the partner countries. Support for the post-launch operation is provided by JAXA and NAOJ (Japan), STFC (U.K.), NASA, ESA, and NSC (Norway). 
The author also would like to thank Enago (www.enago.jp) for the English language review.
\end{acknowledgements}

\bibliography{iida}

\begin{figure}
  \centering
  \subfigure{\includegraphics[width=0.8\columnwidth]{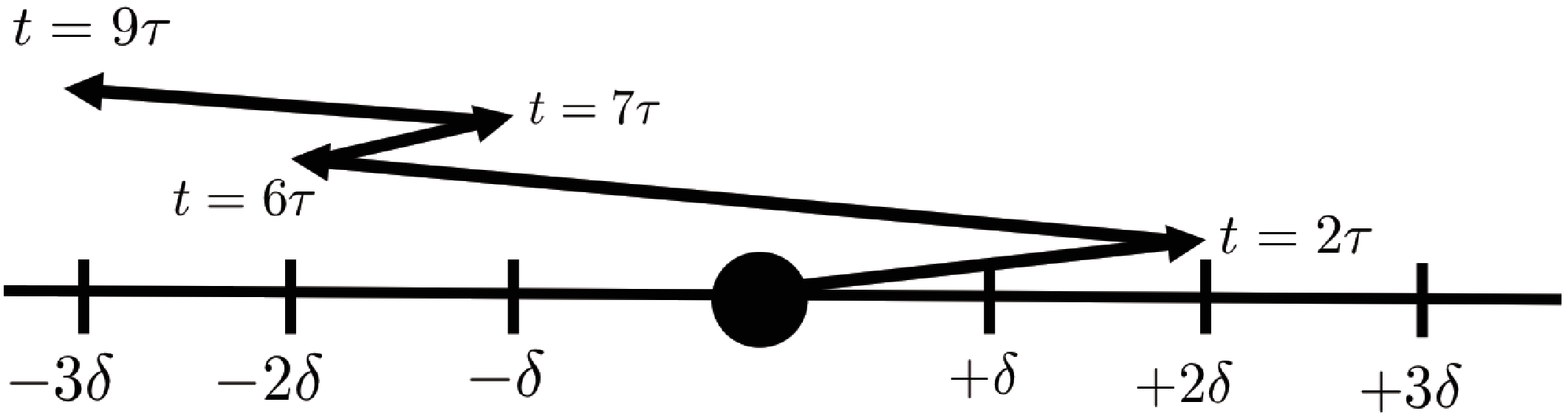}}
  \caption{\small Schematic picture of a 1-D random-walk motion. $\tau$ and $\delta$ are respectively temporal and spatial scales for one walk.}
  \label{f1}
\end{figure}

\begin{figure}
  \centering
  \subfigure{\includegraphics[width=0.8\columnwidth]{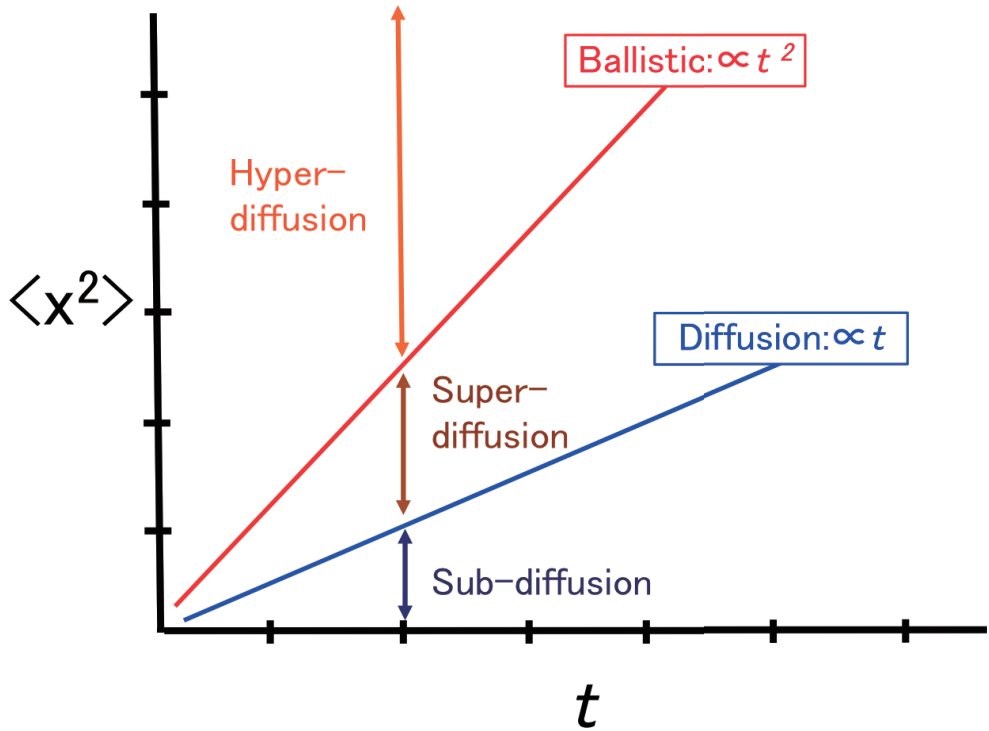}}
  \caption{\small The transport regime on a global scale, based on the relationship between the elapsed time and the squared displacement of each element. Note that the axes are logarithmically scaled.}
  \label{f2}
\end{figure}

\begin{figure}
  \centering
  \subfigure{\includegraphics[width=0.6\columnwidth]{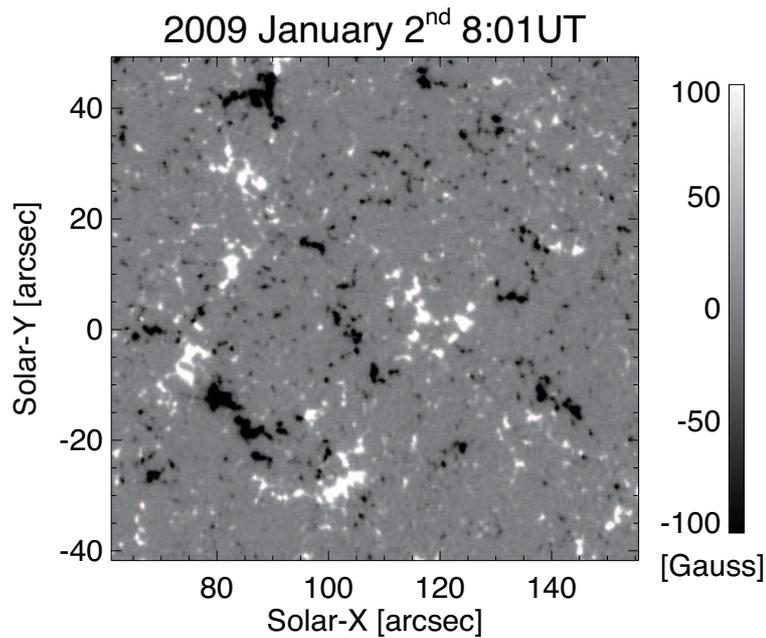}}
  \caption{\small An example of a magnetogram after the preprocessing described in the text.}
  \label{f3}
\end{figure}

\begin{figure}
  \centering
  \subfigure{\includegraphics[width=0.8\columnwidth]{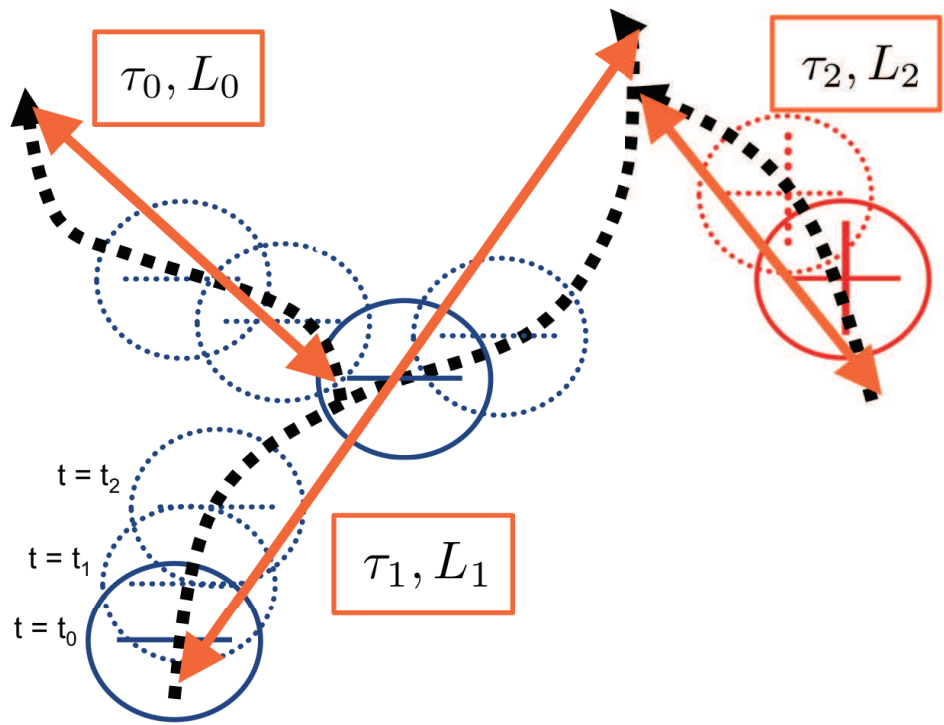}}
  \caption{\small Schematic picture of tracking and squared displacement of the magnetic flux concentration. Three concentrations are tracked. The orange arrows show the displacements for each concentration. Three sets of the elapsed time and the squared displacement are obtained in this example.}
  \label{f4}
\end{figure}

\begin{figure}
  \centering
  \subfigure{\includegraphics[width=0.8\columnwidth]{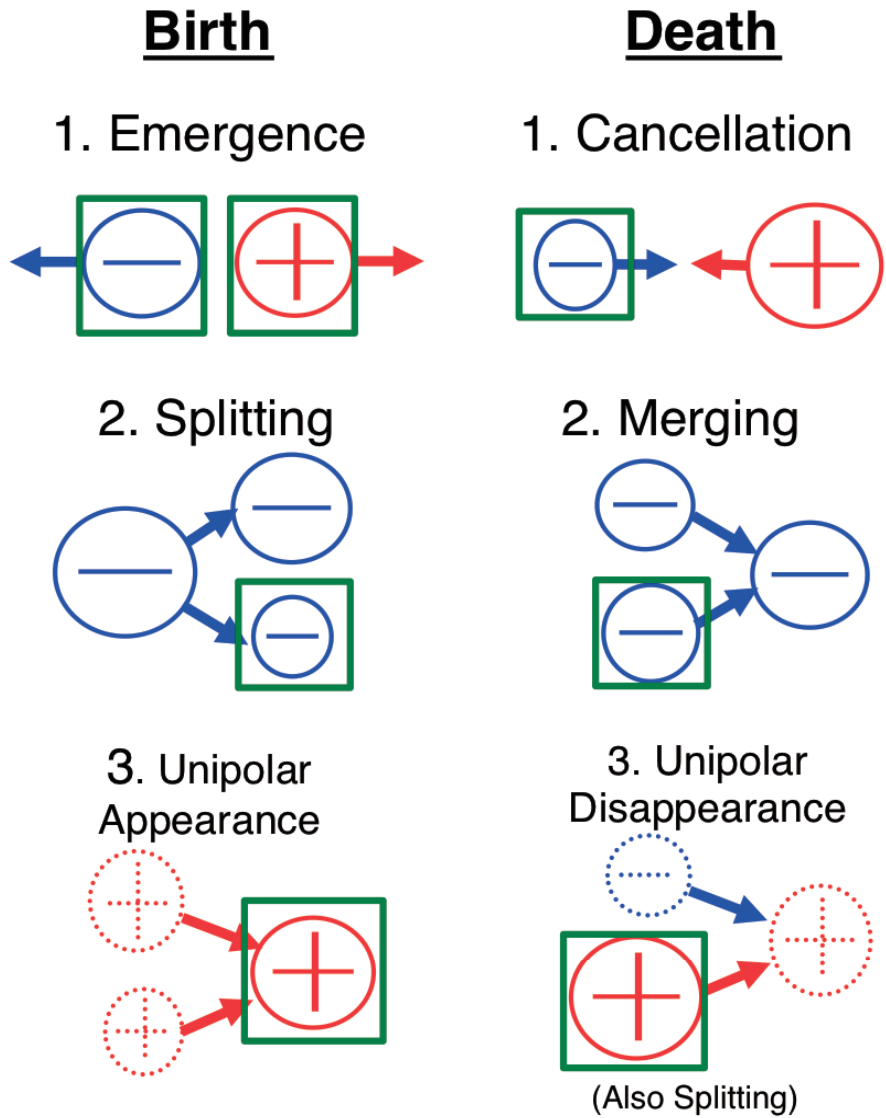}}
  \caption{\small Schematic pictures of births and deaths of magnetic concentrations in our analysis. The left and right columns show the birth and death events, respectively. The concentration surrounded by the green square is that which is produced or dies in each case.}
  \label{f5}
\end{figure}

\begin{figure}
  \centering
  \subfigure{\includegraphics[width=0.9\columnwidth]{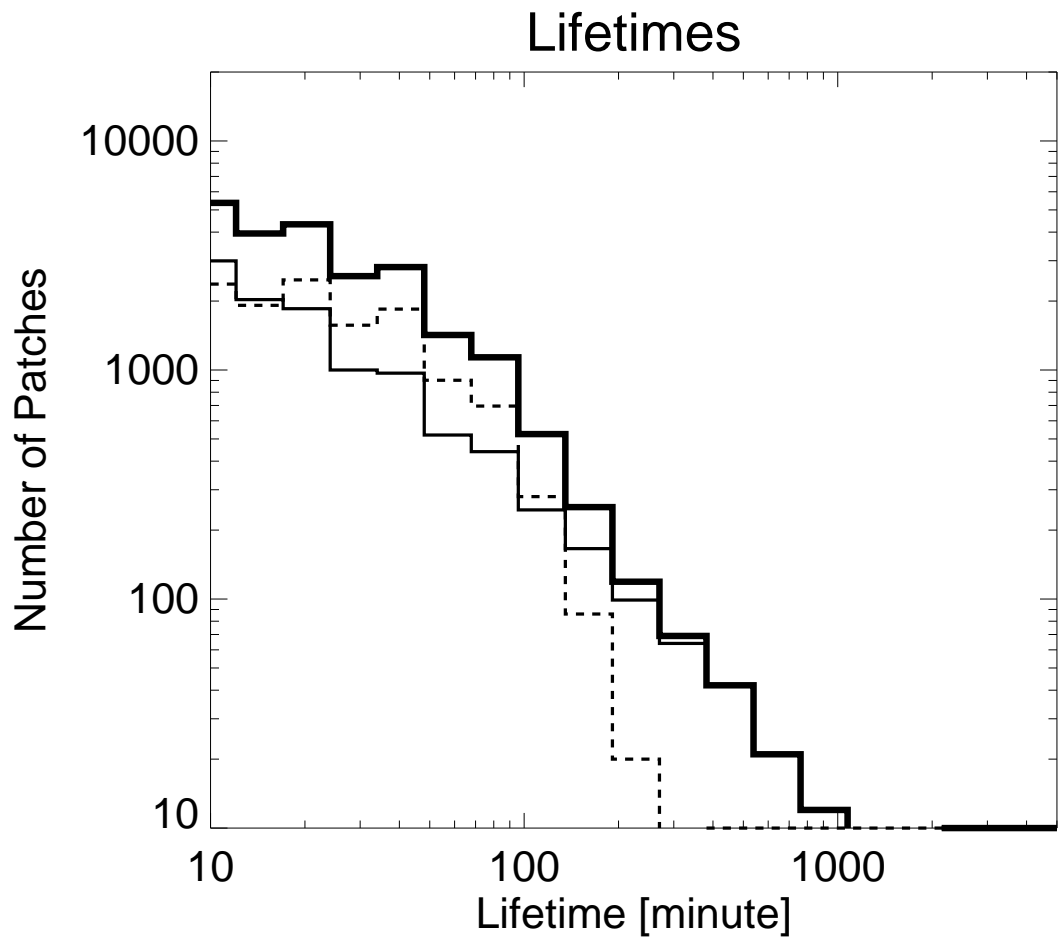}}
  \caption{\small Number histograms of the lifetime. The thin solid and dashed historgrams show those of positive and negative polarities respectively. The thick solid one shows the total of both polarities.}
  \label{f6}
\end{figure}

\begin{figure}
  \centering
  \subfigure{\includegraphics[width=0.9\columnwidth]{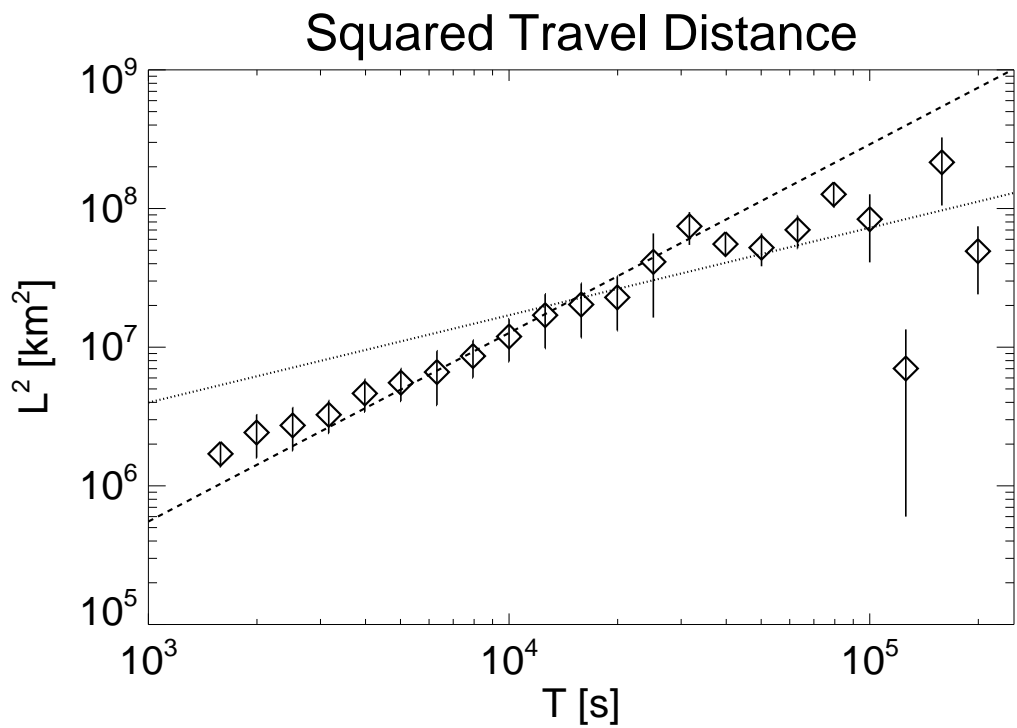}}
  \caption{\small Relationship between the elapsed time and the squared displacement of the patches. Diamonds show the average of the squared displacement and bars shows the one-sigma error in each bin. The dashed and dotted lines show the fitting results below and above the time scale of $2\times10^4$ s, respectively.}
  \label{f7}
\end{figure}

\begin{figure}
  \centering
  \subfigure{\includegraphics[width=0.9\columnwidth]{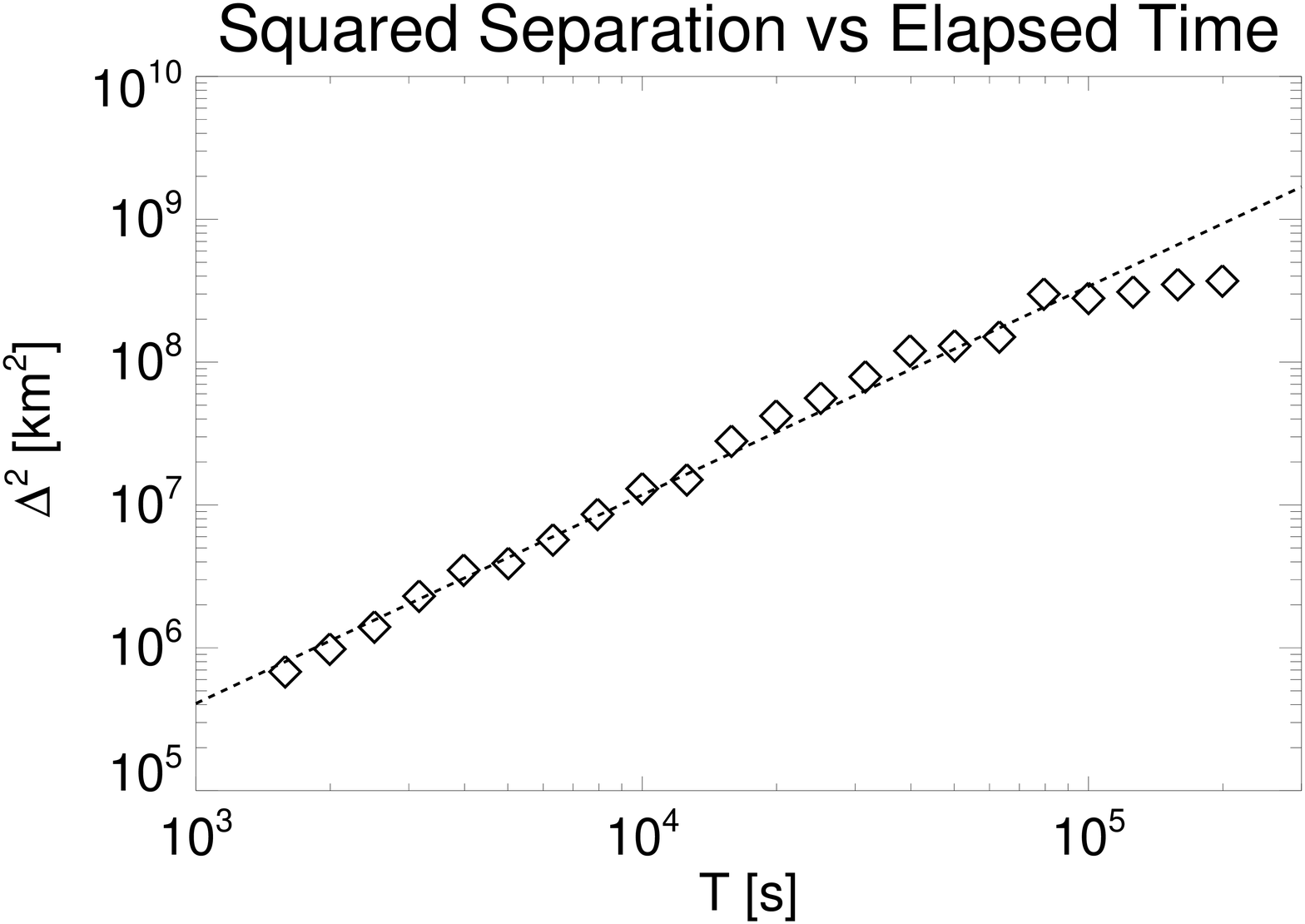}}
  \caption{\small Relationship between the elapsed time and the squared separation distance of the paired patches. 
Diamonds show the average of the squared separation. 
The dashed line corresponds to the power-law fit.}
  \label{f8}
\end{figure}

\begin{figure}
  \centering
  \subfigure{\includegraphics[width=0.9\columnwidth]{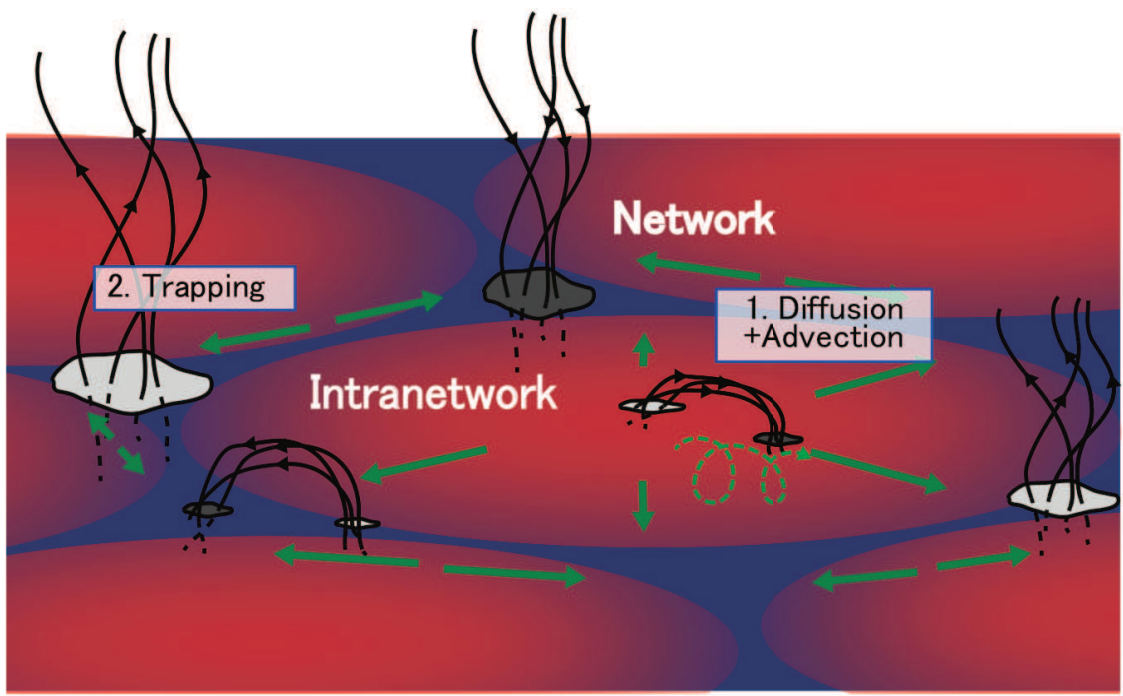}}
  \caption{\small Schematic of the interpretation of the magnetic concentration transport on the network boundary. In and on the network boundary, the motion of the magnetic concentration behaves as a super-diffusion by the network flow and the granular diffusive flow. When it reaches at the conjunction point of the network, it is trapped by the network field and the motion becomes sub-diffusive.}
  \label{f9}
\end{figure}

\end{document}